\documentclass[12pt]{iopart}

\newcommand{\cmmnt}[1]{\ignorespaces}

\usepackage{graphicx}
\usepackage{textcomp}
\usepackage{braket}
\usepackage[mathlines]{lineno}
\usepackage[textsize=tiny]{todonotes}
\usepackage{color}
\begin{document}


\title{30-Fold Increase in Atom-Cavity Coupling Using a Parabolic Ring Cavity }
%
%

\author{$^1$Kevin C. Cox, $^{1,2}$David H. Meyer, $^3$Nathan A. Schine, $^1$Fredrik K. Fatemi, and $^1$Paul D. Kunz}
\address{$^1$US Army Research Lab, $^2$University of Maryland, College Park, $^3$University of Chicago}
 
%
%
%
%
%
 

\date{\today}

\begin{abstract}

Optical cavities are one of the best ways to increase atom-light coupling and will be a key ingredient for future quantum technologies that rely on light-matter interfaces.  We demonstrate that traveling-wave ``ring'' cavities can achieve a greatly reduced mode waist $w$, leading to larger atom-cavity coupling strength, relative to conventional standing-wave cavities for given mirror separation and stability.  Additionally, ring cavities can achieve arbitrary transverse-mode spacing simultaneously with the large mode-waist reductions.  Following these principles, we build a parabolic atom-ring cavity system that achieves strong collective coupling $NC = 15(1)$ between $N=10^3$ Rb atoms and a ring cavity with a single-atom cooperativity $C$ that is a factor of $35(5)$ times greater than what could be achieved with a near-confocal standing-wave cavity with the same mirror separation and finesse. By using parabolic mirrors, we eliminate astigmatism--which can otherwise preclude stable operation--and increase optical access to the atoms.  Cavities based on these principles, with enhanced coupling and large mirror separation, will be particularly useful for achieving strong coupling with ions, Rydberg atoms, or other strongly interacting particles, which often have undesirable interactions with nearby surfaces. 
%
\end{abstract}

%
\maketitle
\newcommand{\daRange}{d_{a,\text{range}}} 
\newcommand{\da}{d_{a}}
\newcommand{\dc}{d_{c}}

Optical cavities increase the coupling between light and matter by an enormous factor (as much as $10^5$) given by  the average number of times light bounces between the mirrors before being transmitted or absorbed, approximately the cavity finesse $F$.  For this reason, optical cavities will continue to be a key ingredient for emerging quantum technologies across a wide range of qubit platforms including neutral atoms \cite{reiserer_cavity-based_2015}, ions \cite{casabone_enhanced_2015}, and solid-state qubits \cite{sipahigil_integrated_2016}. Optical cavity technology continues to advance in a variety of useful directions, for example: extreme stability and finesse \cite{kessler_sub-40-mhz-linewidth_2012}, integrated photonic waveguides \cite{goban_superradiance_2015,tiecke_nanophotonic_2014}, and record atom-cavity coupling achieved with fiber cavities \cite{haas_entangled_2014}.

Within this broad spectrum of cavity technologies, one of the most useful optical cavity designs is still the two-mirror standing-wave (i.e. Fabry-Perot) cavity. While these cavities have recently enabled exquisite quantum control \cite{cox_deterministic_2016,hosten_measurement_2016,mcconnell_entanglement_2015,hacker_photonphoton_2016}, they also have several serious potential drawbacks:  1) The standing-wave results in spatially inhomogeneous light-matter coupling. 2) One cannot simultaneously make the cavity tightly focusing, stable, and single mode. 3) Small-mode cavities limit optical access and readily develop mirror surface charges that perturb strongly interacting qubits, such as ions and Rydberg atoms \cite{steinmetz_stable_2006,steiner_single_2013,casabone_enhanced_2015}.  Ring cavities, made with three or more mirrors, make a directional path for light and naturally provide a solution to the first challenge \cite{culver_collective_2016,kruse_cold_2003,schmidt_dynamical_2014,bao_efficient_2012}.  In this work we solve challenges 2) and 3) by presenting a ring cavity with increased coupling relative to a comparable standing-wave cavity while maintaining stability, large mirror separation, and tunability of transverse modes. 

The large coupling enhancements we discuss can be achieved using either parabolic mirrors, as done here, or a four-mirror ``bowtie'' configuration.  While bowtie enhancement cavities have been used in lasers \cite{nagourney_quantum_2014}, nonlinear optics, and high power laser cavities \cite{carstens_large-mode_2013}, only one recent experiment has explicitly placed atomic qubits inside a ring cavity using a geometry where large coupling enhancements are possible \cite{jia_strongly_2017,ningyuan_observation_2016}.  Here we emphasize the attractive properties of ring cavities, particularly the ability to engineer small mode waists, with the hope that they can be utilized to a greater degree to advance cavity-based quantum technology.

In this paper, we demonstrate strong collective coupling between an ensemble of Rb atoms and a parabolic ring cavity, diagrammed in Fig. 1 a).  The cavity mirrors reside outside the vacuum chamber in a geomtry with one short focusing arm (length $d_1$), containing a cloud of laser-cooled and trapped atoms, and one long nearly-collimated arm (length $d_2$). This ring cavity allows a significantly smaller waist $w_1 = 12(1)$~\textmu m compared to a single-mode standing-wave cavity of the same length, resulting in a factor of 35(5) increase in atom-photon coupling.  Additionally, we demonstrate that the frequency splitting between transverse cavity modes can be tuned arbitrarily while maintaining a small waist, a level of control that is impossible with standing-wave cavities.  Parabolic mirrors eliminate astigmatism and increase optical access, opening the possibility for addressing the atoms at small angles \cite{dupraz_abcd_2015}. The outside-vacuum design has allowed rapid prototype iteration and implementation in our quantum experiments. Despite the moderate finesse $F = 80(3)$ due to traversing the vacuum windows, we achieve a collective cooperativity of $NC = 15(1)$ using only $10^3$ Rb atoms.

\begin{figure}
\begin{centering}
\includegraphics{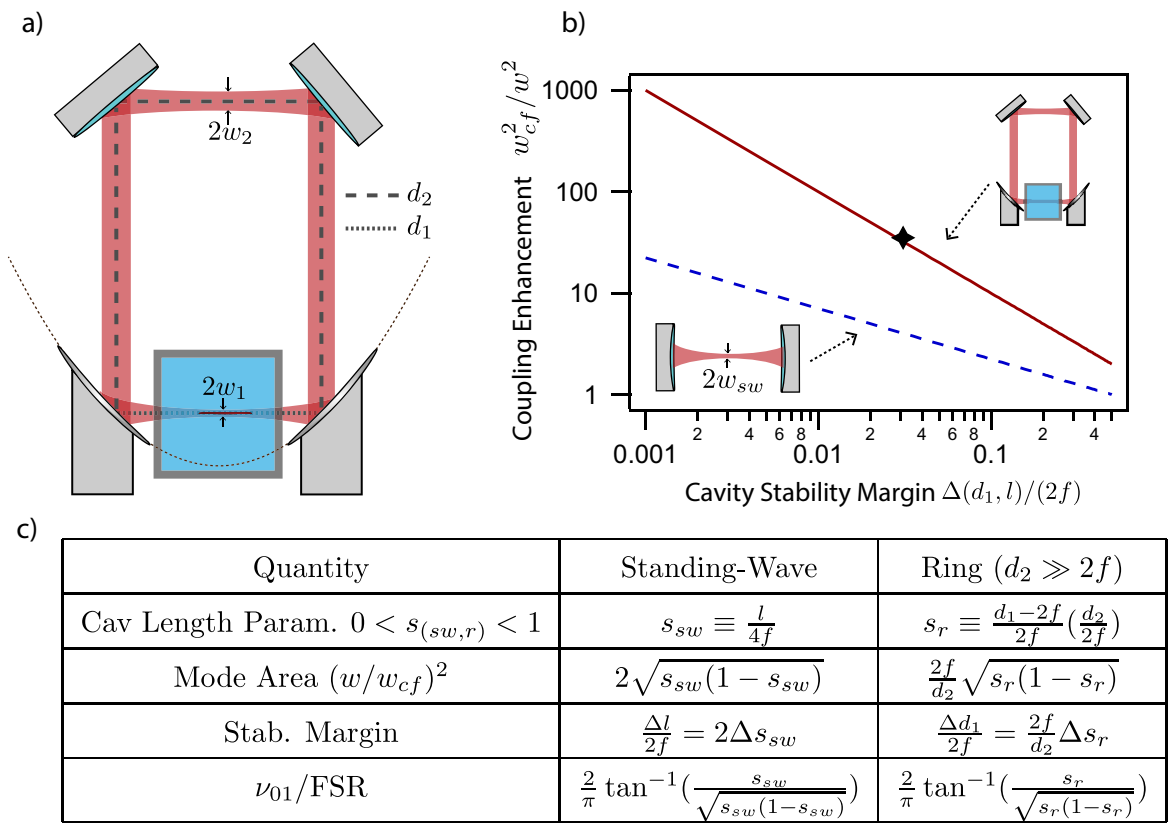}
\caption{ Ring cavity with parabolic mirrors compared to traditional standing-wave cavities. a) Our ring cavity is characterized by two parabolic mirror surfaces separated by distance $d_1 \approx 2f$ where $f$ is the reflected focal length of the parabolic mirrors.  The arm of the cavity without atoms is set to be longer, with length $d_2$.  Rubidium atoms are trapped at the small cavity mode waist $w_1$.  A larger waist $w_2$ is formed directly opposite of the vacuum cell and the atoms. b) The atom-cavity coupling enhancement, given as the waist $w$ (red line for $w_1$ and blue line for $w_{sw}$) relative to the waist of an equivalent near-confocal cavity $w_{cf}$, is plotted versus the cavity stability margin to displacements, $\Delta d_1/f $ for the ring cavity and $\Delta l/f$ for a standing-wave cavity.  Our cavity design is shown as a black star, indicating a couping enhancement of 35(5).  c) Table of analytical results for cavity parameters of standing-wave and ring cavities.  For ring cavities, we have made the approximation $d_2 \gg 2f$.}
\label{fig1}
\end{centering}
\end{figure}

\section{Atom-Cavity Coupling}

Atom cavity experiments are characterized by three rates. $\kappa$ is the cavity linewidth, $\Gamma$ is the atom linewidth due to spontaneous emission into free space, and $g$ is the Jaynes-Cummings parameter that quantifies coupling between the atom and the cavity \cite{reiserer_cavity-based_2015,chen_cavity-aided_2014}.  In many experiments, the important metric for atom-cavity coupling is known as the single atom cooperativity,
\begin{equation}
C = \frac{4 g^2 }{ \kappa \Gamma} = \frac{F \sigma }{\pi A},
\end{equation}
 which can be interpreted as the probability that a single atom emits a resonant excitation as a photon into the cavity mode relative to the probability of emitting into free space \cite{tanji-suzuki_chapter_2011}. For atoms, the cooperativity can be re-written in terms of the atomic scattering cross section $\sigma$, that depends only on the resonant optical wavelength $\sigma \sim \lambda^2$, the cavity finesse $F$, and the cross-sectional area $A$ of the optical mode at the location of the atom.  $C$ is independent of cavity length, and can only be increased by improving the cavity finesse or decreasing the area of the optical mode.  Increasing the cavity finesse can often be undesirable because the bandwidth of light exchange between the cavity and the input/output modes (i.e. $\kappa$) is decreased.  Here we show a route to improving $C$ by decreasing the mode area, proportional to mode waist squared $A \sim w^2$, thus increasing the electric field strength for the atoms inside.

\section{Ring Cavities vs. Standing-Wave Cavities}

The mode waist $w$ and stability of standing-wave cavities can be found in a number of textbooks \cite{nagourney_quantum_2014, siegman_lasers_1986}.  Standing-wave cavities are most stable near the confocal regime, where the cavity length, $l$, is equal to twice the mirror focal length, $f$.  In this situation, the mode waist is $w_{cf} \equiv \sqrt{ f \lambda/\pi}$.  $w_{cf}$ is fixed by the fact that confocal standing-wave cavities have a Rayleigh range  equal to half the cavity length $l$.  The only way to decrease this waist is to decrease $f$. Since creating extremely small focal length mirrors can be challenging, and one often does not want mirrors close to the qubits, the goal here is to engineer cavities with a significant waist reduction $w/w_{cf} \ll1$.

To achieve small cavity waists with standing-wave cavities, one must go to the edge of stability, the ``concentric" ($l = 4f$) or ``planar'' regime ($l\ll f$).  In both of these regimes the cavity waist asymptotically approaches zero; however, there are two negative side effects.  First, the cavity becomes highly sensitive to misalignment and length change.  Second, the transverse modes, defined by Hermite-Gaussian or Laguerre-Gaussian functions, become degenerate, so the spatial profile of the light inside the cavity cannot be controlled \cite{siegman_lasers_1986}.

Consider instead a ring cavity as in Fig. \ref{fig1} a).  Using Gaussian ABCD matrices, one can derive the cavity waist by solving the cavity's self-consistency condition (mode matches itself after one cavity round trip).  Full analytical results can be found in Ref. \cite{nagourney_quantum_2014}. We constrain ourselves to the regime $d_2\gg 2f$ and $d_1 \approx 2f$, such that a small waist appears in the short $d_1$ arm.  

In Fig. \ref{fig1} c) we give analytic expressions for dimensionless versions of the standing-wave and ring cavity parameters (labeled with subscripts $sw$ and $r$ respectively).  The cavity length is re-defined as the dimensionless parameter $s_{(sw,r)}$ which spans from 0 to 1 over the stability ``island" (see Fig. 2).  The waists are all normalized to the characteristic waist of a confocal cavity $w_{cf}$.  We define a stability margin ($\Delta d_1$ for ring cavity or $\Delta l$ for standing wave) as the distance the cavity length can be changed before the cavity becomes unstable.  We also include an expression for the normal mode splitting, $\nu_{01}$, normalized to the cavity free spectral range (FSR).  Fig. \ref{fig1} c) shows that the waist, stability, and mode splitting of standing-wave cavities are all determined solely by the parameter $s_{sw}$.  In the ring cavity expressions, on the other hand, the additional factor of $2f/d_2 \approx d_1/d_2$ in the waist and stability allows independent tuning.  For this reason, ring cavities can achieve small mode waists while maintaining both a reasonable stability margin and arbitrary transverse mode spacing.  

To highlight the increased atom-cavity coupling of the ring cavity, in Fig. \ref{fig1} b) we plot the coupling enhancement given by a ring cavity (red line), relative to a confocal cavity with the same focal length mirrors, as a function of the fractional length displacement over which the cavity remains stable $\Delta d_1$. Our cavity achieves a waist $12(1)$~\textmu m, and has a single-atom cooperativity that is $35(5)$ times higher than a confocal (or nearly-confocal) cavity. For comparison, the waist that can be achieved by moving a standing wave cavity close to the concentric regime is plotted as a blue-dashed line. Given comparable stability, concentric cavities can only achieve a cooperativity enhancement of less than $5$ relative to confocal cavities.  Furthermore, ring cavities have an improved scaling of mode area versus stability margin, $w_1^2 \propto \Delta d_1$, while for standing-wave cavities the area only scales as the square root of the stability margin $w_{sw}^2 \propto  \sqrt{\Delta l}$. 

For our ring cavity, the waist $w_2$ becomes larger, and more collimated, as $d_2$ is increased.  The large beam width at the location of the parabolic mirrors leads to a tight waist $w_1$.  This principle may be extended by utilizing two additional \textit{convex} mirrors placed about the large waist, $w_2$, as in \cite{jia_strongly_2017}. Such mirrors expand the beam more rapidly than free-space propagation; the larger mode size on the concave mirrors then results in a tighter focus. This allows the long arm of the cavity to be shortened while maintaining a small waist, which is important for experiments that seek $g \gg \Gamma$.  We emphasize that a long cavity length $d_2$ is not fundamentally required to achieve the large waist reductions reported here.

\begin{figure}[htb]
\begin{centering}
\includegraphics{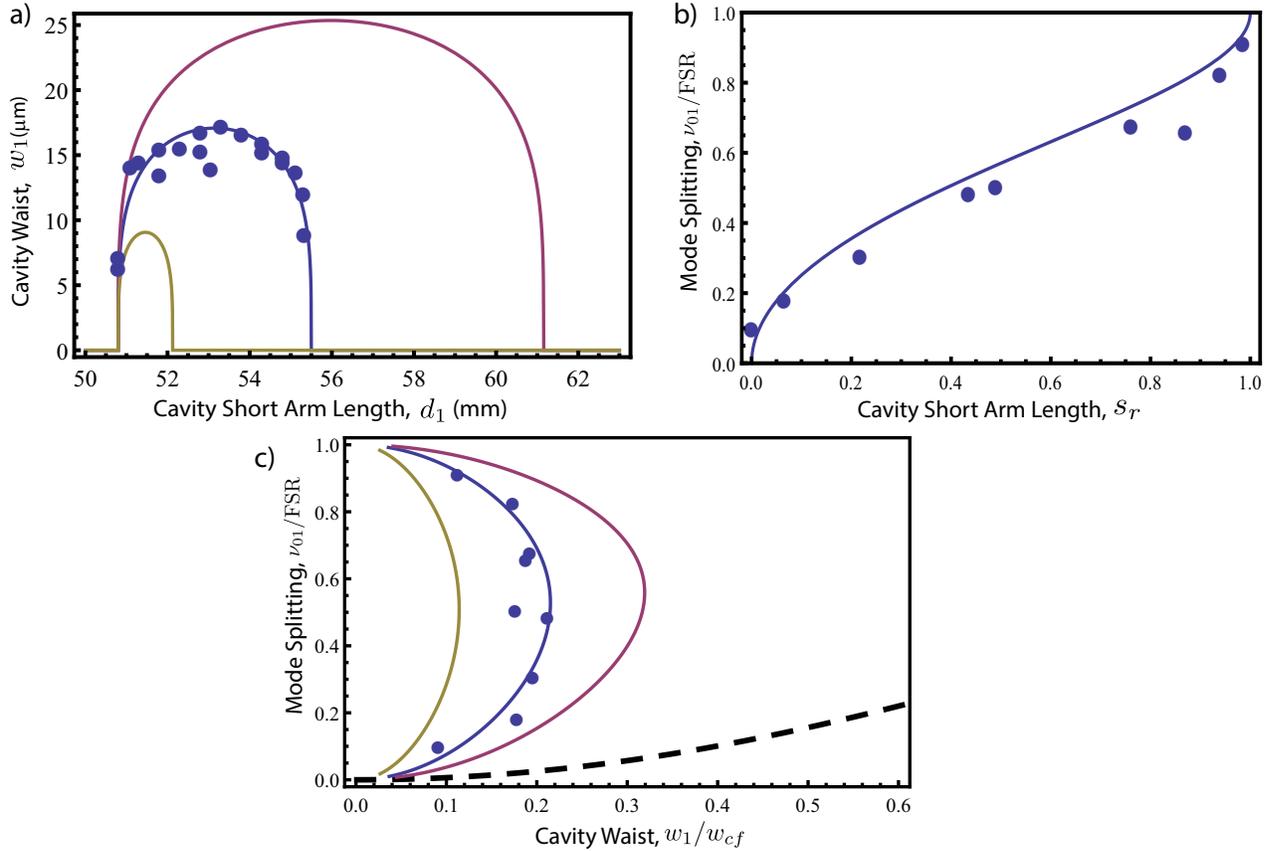}
\caption{Properties of ring cavities.  a) The stability region of the optical cavity is shown as a function of short arm distance $d_1$.  As the long arm of the cavity becomes longer, the stability island shrinks leading to a smaller mode waist ($d_2 = 30$~cm in red, $d_2 = 60$~cm in blue, $d_2 = 200$~cm in gold).  Experimental measurements of the mode waist taken with a parabolic ring cavity with $d_2 = 60$~cm are shown. b) Mode splitting of the ring cavity.  By modifying $d_1$ over the stability island the splitting between transverse modes can be tuned arbitrarily between zero and a full free spectral range. c) Mode splitting versus waist.  Possible transverse mode splittings are plotted versus cavity waist for concentric standing-wave cavities (black dashed line) and ring cavities.  Our standalone cavity, with $d_2 = 11 f$ is shown theoretically in blue with experimental data points.  Ring cavities allow arbitrary transverse mode splitting (example curves shown for $d_2/(2f) = 40$ (gold) and $d_2/(2f) = 6$ (red)) for any cavity waist, allowing access to any point on the plot. Standing-wave cavities quickly become degenerate and are limited to points on the black dashed line of this plot. }
\end{centering}
\label{ringCavData}
\end{figure}
\section{Standalone Cavity}

We constructed a standalone optical ring cavity (with no atoms inside) of the form shown in Fig. \ref{fig1} to demonstrate that ring cavities can realize the expected reduction in mode waist while maintaining a large stability margin and control over the transverse mode splitting.  The cavity was formed by one fixed parabolic mirror with the other on a translation stage. For this cavity we used gold coated parabolic mirrors with $f = 2.54$~cm that, along with a 5\% transmission input coupler limited the finesse to $F = 50$. The long arm cavity length was $d_2 = 60$~cm, and the short arm could be tuned over the entire stability range, approximately 51 to 56~mm, with a translation stage.  Standard kinematic mirror mounts were sufficiently stable given our moderate finesse which made cavity alignment relatively easy. A pellicle pick-off, also on a translation stage, was placed within the short arm for imaging the beam profile using a microscope objective with 26x magnification. 

Figure 2 shows the measured and predicted properties of ring cavities.  The cavity waist is plotted in part a) as a function of the short cavity arm length $d_1$ for three different values of the long arm length $d_2$.  The stability island shrinks as $d_2$ becomes longer, corresponding to a larger, more collimated beam in the long arm and a smaller waist in the short arm.  We also show measured values (blue points) from our standalone ring cavity.  We achieve high quality cavity modes with no measurable astigmatism and waists down to $7$~\textmu m.  Such observations would be difficult, even with a bowtie cavity, where remaining astigmatism from angled reflections places a limit on the smallest achievable waist \cite{nagourney_quantum_2014}.  We expect that our ring cavities, on the other hand, can approach the diffraction limit given by the numerical aperture of the mirrors, but this is outside the scope of this current work, where only strong collective coupling is desired.

In Fig. 2 b), we measure and plot the frequency splitting $\nu_{01}$ between the nearest observed $LG_{10}$ and $LG_{00}$ modes as we move across the stability island shown in Fig. 2 a).  The mode splitting $\nu_{01}$ can be varied over the entire free spectral range as one tunes over the stability island.  The tunability of $\nu_{01}$ arises from having a second cavity arm ($d_2$) with a variable Gouy phase shift, giving complete control of the mode splitting for a given ring cavity while maintaining a small waist. This ability to arbitrarily tune $\nu_{01}$ is important for atom-light interfaces which usually desire higher order cavity modes to be far off resonance from the fundamental mode.  On the other hand, our ring cavity design is also desirable for the recent cavity experiments that have explicitly engineered mode degeneracy to simulate many-body quantum systems \cite{schine_synthetic_2016,sommer_engineering_2016,kollar_adjustable-length_2015}.

In Fig. 2 c) we show the mode splitting $\nu_{01}$, as a function of cavity waist.  For a standing-wave cavity, near the concentric point, the mode splitting is constrained to lie on the thick black dashed line, a small fraction of the FSR. However, our ring cavity scheme gives independent control of the mode splitting and the cavity waist.  By changing the long arm length $d_2$, one can access any point on the plot (theoretical curves shown for several values of $d_2/(2f)$).  Our standalone cavity has a maximum waist of approximately $15$~\textmu m, and to achieve this in the concentric standing-wave configuration would mean that the mode splitting would be less than the current cavity linewidth.

\section{Cavity with Atoms}

Finally, we combine the parabolic ring cavity with a laser-cooled atom experiment to demonstrate strong coupling between the cavity and the atomic ensemble.  The parabolic ring cavity used in this setup is similar to the standalone version, but uses higher reflectivity $R = 98.2\%$ dielectric parabolic mirrors with focal length $f= 2$~cm, and a long arm of $d_2 = 56$~cm. We are able to reproducibly align the cavity using standard kinematic mirror mounts, a result of the intrinsic robustness of this geometry and the low finesse.  The vacuum chamber is anti-reflection coated and has a measured total transmission of $97.5\%$, which is limited by a combination of imperfect coating and rubidium deposits in the glass. Rubidium vapor is cooled in a 3D magneto-optical-trap (MOT), and approximately $N = 10^3$ atoms are then loaded into a 1-D running-wave optical dipole trap created by a far-off resonant cavity mode at $785$~nm.  The atoms are laser-cooled with polarization gradient cooling to approximately $20$~\textmu K.  One cavity mirror is mounted to a piezo and the cavity resonance frequency is stabilized to the trapping beam via a Pound-Drever-Hall lock.  We use two levels in rubidium, as shown in Fig. 3.   The atoms are probed using the $780$~nm optical transition between a hyperfine ground state $\ket{\uparrow} = \ket{5_{s,1/2},F=2}$ and an excited state $\ket{e} = \ket{5_{p,3/2}, F=3}$.

To demonstrate strong collective coupling between the atoms and the cavity, we tune the cavity frequency $\omega_c$ on resonance with the $\ket{\uparrow}\rightarrow\ket{e}$ transition.  The coupling between the atoms and cavity leads to a normal mode splitting, often referred to as a vacuum Rabi splitting (VRS) $\Omega$ \cite{chen_cavity-aided_2014}.  The modes are split by an amount equal to $\Omega = 2 g\sqrt{N}$, collectively enhanced by the atom number $N$.  When $NC$ is large, the width of the features $\kappa'$ is the geometric mean of the atom and cavity linewidth, $\kappa' = \sqrt{\kappa \Gamma}$, so that the magnitude of the VRS relative to the full width at half maximum of the lines is a direct measurement of the collective cooperativity $NC$ \cite{chen_cavity-aided_2014}

Figure 3 shows a plot of the measured cavity mode with and without atoms.  We measure the fractional probe laser power reflected, $R$, as the laser frequency detuning $\Delta_p$ is swept over the atom-cavity resonance.  The power is kept low so we do not saturate or optically pump the atoms.  Without atoms, we observe a Lorentzian reflection dip corresponding to the empty cavity linewidth, measured to be $\kappa = 6.4(1)$~MHz.  With atoms, we observe a well-resolved VRS with splitting $\Omega =25(1)$~MHz, which corresponds to a collective cooperativity $NC = \Omega^2/(\kappa \Gamma) = 15(1)$.  This strong cooperativity is the key figure of merit for many quantum technologies and shows that this system can be useful for, among other things, quantum communication and entanglement generation \cite{reiserer_cavity-based_2015,chen_cavity-aided_2014}.

\begin{figure}[htb]
\begin{centering}
\includegraphics{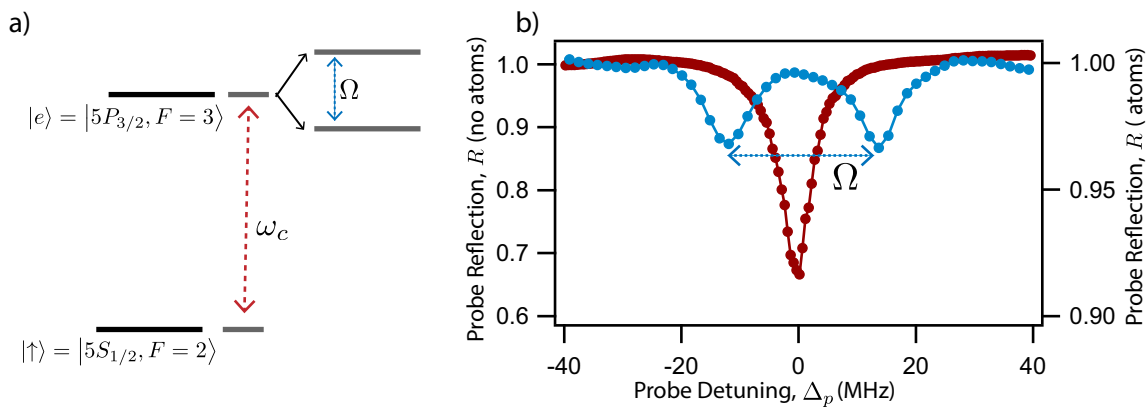}
\caption{(Color online) Vacuum Rabi splitting.  a) Level diagram.  The optical cavity is tuned on resonance with the $780$~nm D2 transition, leading to a normal mode splitting (VRS) of magnitude $\Omega$.  b) Probe reflection without atoms (red) shows a single reflection dip associated with the optical cavity, with linewidth $\kappa = \times 6.4(1)$~MHz. Probe reflection with approximately $1500$ atoms (blue) shows two modes, separated by $\Omega = 25(1)$~MHz, confirming strong collective coupling with the atoms.}
\end{centering}
\label{fig3}

\end{figure}

The single-particle cooperativity $C$ is calculated from the finesse of the optical cavity and the cavity waist $w_1$.  $w_1 = 12(1)$ \textmu m is confirmed by fluorescence imaging of the cavity mode with the vacuum cell background Rb vapor.  The measured finesse is $80(3)$, primarily limited by transmission of the input coupler ($T = 0.02$), loss in the vacuum cell ($T = 0.975$), and the parabolic mirrors ($R = 0.982$), leading to a peak single-atom cooperativity $C = 0.012(2)$  One usually desires to have an optical cavity limited by transmission rather than loss since losses result in an effective quantum efficiency for cavity readout of $q \equiv T/(L+T)$.  Given our current cavity losses we achieve $q = 25\%$. However, these losses can be reduced with a high-quality coated, 2D-MOT loaded, vacuum cell with lower rubidium content and by further improving the parabolic mirror reflectivity. This would raise the quantum efficiency to over $90\%$ while maintaining the demonstrated strong coupling.
\section{Conclusion}

In the future, we hope to utilize this setup for a state-of-the-art atomic quantum memory.  For quantum memories, ring cavities are necessary to distinguish between forward and backward scattered photons, which give significantly different diffusion-limited spin-wave lifetimes \cite{simon_interfacing_2007}.  The ring cavity also gives a large amount of filtering between the optical trap and signal photons.  Lastly, our parabolic design provides a large mirror surface which can be used to reflect other beams onto the atoms, giving excellent optical access for multiple write beams and allowing for the possibility of significant spatial multiplexing \cite{sangouard_quantum_2011}.  A triangular ring cavity has already been used in a long lifetime atomic quantum memory \cite{bao_efficient_2012}, but the techniques of this paper were not employed and strong coupling was not demonstrated.  

In addition to increased atom-cavity coupling, ring cavities open up additional opportunities that are significant for future quantum applications.  With four or more mirrors, for example, a slight non-planarity in the round-trip path introduces a geometric rotation of the polarization vector that may entirely overcome the linear birefringence of the cavity mirrors. This would result in spectrally resolved circularly polarized eigenmodes that enable intra-cavity optical pumping of an atomic gas. Recently, these same geometric cavity mode rotations have been used to introduce a strong synthetic magnetic field for photons \cite{schine_synthetic_2016,sommer_engineering_2016}.

We use this cavity to achieve strong collective coupling with an outside-vacuum cavity, but another possible future direction could be to use a high-finesse, in-vacuum version to achieve single-atom strong coupling.  This approach may be an easier alternative compared to a number of experiments which have sought this limit using extremely small focal length mirrors \cite{northup_ion-cavity_2015} or concentric cavities \cite{durak_diffraction-limited_2014}.  Large mirror separation is especially critical for ions or Rydberg atoms which strongly interact with nearby dielectric surfaces.  Although high-finesse parabolic mirrors are not commonly available, there should be no inherent difficulty in achieving finesse greater than 1000.  A quick calculation suggests that using a cavity with $f = 1$~cm and $d_2 = 25$~cm could reach the single-atom strong cooperativity regime with $C>10$ with a cavity linewidth of $\kappa \sim 0.1$~MHz and a finesse of 10,000. Overall, we believe that ring cavities, currently under-utilized, will play a more important role in the future of quantum technologies.

\section{Acknowledgements}
We thank Michael Foss-Feig, Jon Simon and James Thompson for useful discussions.  This work is supported by the US Army Research laboratory and the Quantum Science and Engineering Program from the US Office of the Secretary of Defense. David Meyer acknowledges support from Oak Ridge Associated Universities.\\
\\

\bibliographystyle{unsrt}
\bibliography{RingCavityPaper}

\end{document}